\documentclass[runningheads]{llncs}
\usepackage{amssymb}
\usepackage{graphicx}
\usepackage{tikz}
\usepackage{amsmath}
\usepackage{mathrsfs}
\usepackage{graphicx}
\usetikzlibrary{automata, positioning, arrows}
\usepackage{mathtools}
\usepackage{xcolor}
\usepackage{stackengine}

\begin{document}
\title{Generalized Linear One-Way Jumping Finite Automata}

\author{Ujjwal Kumar Mishra \and
Kalpana Mahalingam
\and
Rama Raghavan}
\authorrunning{ukmishra et al.}

\institute{Department of Mathematics, Indian Institute of Technology Madras,\\ Chennai-600036, India.\\
\email{ma16d030@smail.iitm.ac.in}\\
\email{\{kmahalingam,ramar\}@iitm.ac.in}}
\maketitle            
\begin{abstract}
A new discontinuous model of computation called one-way jumping finite automata was defined by H. Chigahara et. al. This model was a restricted version of the model jumping finite automata. These automata read an input symbol-by-symbol and jump only in one direction. A generalized linear one-way jumping finite automaton makes jumps after deleting a substring of an input string and then changes its state. These automata can make sequence of jumps in only one direction on an input string either from left to right or from right to left. We show that newly defined model is powerful than its original counterpart. We define and compare the variants, generalized right linear one-way jumping finite automata and generalized left linear one-way jumping finite automata. We also compare the newly defined models with Chomsky hierarchy. Finally, we explore closure properties of the model.     

\end{abstract}



\section{Introduction}
First discontinuous model of computation called general jumping finite automata $\it{(GJFA)}$ was introduced in \cite{JFA2012} by Meduna et. al. These automata read the given input in a discontinuous manner. The automata can jump in either direction to read the input. In \cite{basicpropertiesJFA}, the author solved questions related to closure properties of the model $\it{GJFA}$ which were left open in \cite{JFA2012}. The author showed that universality, equivalence and inclusion are undecidable for $\it{GJFA}$, in \cite{trodip}. Some decision problems of the model jumping finite automata ($\it{JFA}$), which is a restricted version of $\it{GJFA}$, had been discussed in \cite{oscadojfa}. Results related to the complexity of the model $\it{JFA}$ were discussed in \cite{cacrojfa} and \cite{oscadojfa}. Following \cite{JFA2012}, several other jumping transition models have been defined and studied in \cite{opvojfa,owjfa,ndrowjfa,eoawjm,odjfaatcp,WKJFA,J53WKA}.

The model which is of our interest is one-way jumping finite automata ($\it{OWJFA}$), defined in \cite{owjfa}. The model $\it{OWJFA}$ is defined by giving a restriction on jumping behaviour of the model $\it{JFA}$. There are two variants of $\it{OWJFA}$: right one-way jumping finite automata ($\it{ROWJFA}$) and left one-way jumping finite automata ($\it{LOWJFA}$). $\it{ROWJFA}$ starts processing an input string from the leftmost symbol of the input whereas $\it{LOWJFA}$ starts processing an input string from the rightmost symbol. The models can
only jump over symbols which they cannot process in its current state. Some properties of $\it{ROWJFA}$ are given in \cite{porowjfa}.  The decision problems of the model $\it{ROWJFA}$ are discussed in \cite{dorowjfa}. Nondeterministic variant of the model $\it{ROWJFA}$ is defined and studied in \cite{ndrowjfa}.

$\it{OWJFA}$ processes one symbol at a time, i.e., it go from one state to another by reading a symbol. We define a new jumping transition model called generalized linear one-way jumping finite automata ($\it{GLOWJFA}$). These automata can process a subword in a state. We show that this generalization increases the power of $\it{OWJFA}.$ Similar to $\it{OWJFA}$, we define two types of $\it{GLOWJFA}$: generalized right linear one-way jumping finite automata ($\it{GRLOWJFA}$) and generalized left linear one-way jumping finite automata ($\it{GLLOWJFA}$). $\it{GRLOWJFA}$ delete an input starting from the leftmost end of the input. They can jump over a part of the input if they cannot read it. These automata  make jumps from left to right. But, if at any stage, the present state cannot read any subwords to the right of it, then the automaton makes a jump from right to left and jumps before the leftmost symbol of the input tape. Similarly, $\it{GLLOWJFA}$ process an input from right to left, starting from the rightmost end of the input. They can jump over a part of the input if they cannot read. These automata jump from right to left. But, if at any stage, the present state cannot read any subwords to the left of it, then the automaton makes a jump from left to right and jumps before the rightmost symbol of the input tape.  

In this paper, we compare the models $\it{GLOWJFA}$ and $\it{OWJFA}$. We also compare $\it{GRLOWJFA}$ and $\it{GLLOWJFA}$. The language classes of $\it{GRLOWJFA}$ and $\it{GLLOWJFA}$ are compared with the language classes of Chomsky hierarchy. Closure properties of the language classes of $\it{GRLOWJFA}$ and $\it{GLLOWJFA}$ are explored.

This paper is organised as follows: In Section \ref{prelim}, we give some basic notion and notation. We also recall the definitions of $\it{ROWJFA}$, $\it{LOWJFA}$ and give an example of $\it{ROWJFA}$, in this section. The models $\it{GRLOWJFA}$ and $\it{GLLOWJFA}$ are defined in Section \ref{glowjfa}. We give examples for our new definitions. In Section \ref{glowjowj}, we compare $\it{GRLOWJFA}$ with $\it{ROWJFA}$ and $\it{GLLOWJFA}$ with $\it{LOWJFA}$. The models $\it{GRLOWJFA}$ and $\it{GLLOWJFA}$ are compared in Section \ref{rightleft}. In Section \ref{chomsky}, the language classes of the newly introduced models are compared with the language classes of Chomsky hierarchy. Finally, we discuss closure properties of the newly introduced models, in Section \ref{closure}. We end the paper with few concluding remarks.

\section{Preliminaries}\label{prelim} 
In this section, we recall some basic notations and definitions. An alphabet set is a finite non-empty set $\Sigma$. The elements of $\Sigma$ are called letters or symbols. A word or string $w=a_1a_2 \cdots a_n$ is a finite sequence of symbols, where $a_i \in \Sigma$ for $1 \leq i \leq n$. The reverse of $w$ is obtained by writing the symbols of $w$ in reverse order and denoted by $w^R$, hence $w^R=a_n \cdots a_2a_1$. By $\Sigma^*$, we denote the set of all words over the alphabet $\Sigma$ and by $\lambda$, the empty word. A language $L$ is a subset of $\Sigma^*$ and $L^\mathrm{C} = \Sigma^* \setminus L$ denotes the complement of $L$. The symbol $\emptyset$ represents the empty language or empty set. For an arbitrary word $w \in \Sigma^*$, we denote its length or the number of letters in it by $|w|$. For a letter $a \in \Sigma$, $| w |_a$ denotes the number of occurrences of $a$ in $w$. Note that, $\Sigma^+=\Sigma^* \setminus \{ \lambda \}$ and $|\lambda|=0$. A word $y \in \Sigma^*$ is a subword or substring of a word $w \in \Sigma^*$ if there exist words $x, z \in \Sigma^*$ such that $w=xyz$. If $w=uv$,
then the words $u \in \Sigma^*$ and $v \in \Sigma^*$ are said to be a prefix and a suffix of $w$, respectively. Two sets $A$ and $B$ are comparable if $A \subseteq B$ or $B \subseteq A$. For a finite set $A$, $|A|$ denotes the number of elements in $A$. For two sets $A$ and $B$, if $A$ is a proper subset of $B$, then we use the notation $A \subset B$. For definitions of other basic language operations (like union, intersection etc.), the reader is referred to \cite{book}.

We now recall the definitions of right and left one-way jumping finite automata \cite{owjfa}.
A right  one-way jumping finite automaton $(\it{ROWJFA})$ is a tuple $\mathcal{A}=(\Sigma, Q, q_0, F, R)$, where $\Sigma$ is an alphabet set, $Q$ is a finite set of states, $q_0$ is a starting state, $F \subseteq Q$ is a set of final states and $R \subseteq Q \times \Sigma \times Q$ is a set of rules, where for a state $p \in Q$ and a symbol $a \in \Sigma$, there is at most one $q \in Q$ such that $(p,a,q) \in R$. By a rule $(p,a,q) \in R$, we mean that the automaton goes to the state $q$ from the state $p$ after deleting the symbol `$a$'.  For $p \in Q$, we set $$\Sigma_p=\Sigma_{R,p}=\{b \in \Sigma 
: (p,b,q) \in R ~ for~ some~ q \in Q\}.$$ A configuration of the right one-way jumping automaton $\mathcal{A}$ is a string of $Q\Sigma^*$. The right one-way jumping relation, denoted as $\circlearrowright_{\mathcal{A}}$, over $Q\Sigma^*$ is defined as follows. Let $(p,a,q) \in R, x \in (\Sigma \setminus \Sigma_p)^*$ and $y \in \Sigma^*$. Then, the $\it{ROWJFA}$ $\mathcal{A}$ makes a jump from the configuration $pxay$ to the configuration $qyx$, written as $pxay \circlearrowright _{\mathcal{A}} qyx$ or just $pxay \circlearrowright  qyx$ if it is clear which $\it{ROWJFA}$ is being referred. Let $\circlearrowright^+$ and $\circlearrowright^*$ denote the transitive and reflexive-transitive closure of $\circlearrowright$, respectively. The language accepted by $\mathcal{A}$ is $$L_R(\mathcal{A})=\{w \in \Sigma^* : q_0w \circlearrowright^* q_f ~ for~some ~ q_f \in F\}.$$

A left one-way jumping finite automaton $(\it{LOWJFA})$ is similar to that of a $\it{ROWJFA}$ except that a configuration of $\it{LOWJFA}$ is a string of $\Sigma^*Q$. By a rule $(q,a,p) \in R$, we mean that the automaton goes to the state $q$ from the state $p$ after deleting the symbol `$a$'. For $(q,a,p) \in R, x \in (\Sigma \setminus \Sigma_p)^*$ and $y \in \Sigma^*$ the  $\it{LOWJFA}$ makes a jump from the configuration $yaxp$ to the configuration $xyq$, written as $xyq \prescript{}{\mathcal{A}}{\circlearrowleft}$ $yaxp$ or just $xyq \circlearrowleft yaxp $ if it is clear which $\it{LOWJFA}$ is being referred.  The language accepted by $\mathcal{A}$ is $$L_L(\mathcal{A})=\{w \in \Sigma^* : q_f \prescript{*}{}{\circlearrowleft}~ wq_0 ~ for~some ~ q_f \in F\}.$$

\begin{example}

Consider the $\it{ROWJFA}$ $\mathcal{A}=(\{a,b\},\{q_0,q_1,q_2,q_3\},q_0,\{q_2\},R)$, where $R=\{(q_0,b, q_1),(q_0,a, q_2),(q_2,a,q_3), (q_3,b,q_2)\}$. Here, $\Sigma_{q_0}=\{a,b\}$, $\Sigma_{q_1}=\emptyset$, $\Sigma_{q_2}=\{a\},\Sigma_{q_3}=\{b\}$.
Since $\Sigma_{q_0}=\{a,b\}$,  the automaton cannot jump over any symbol $`a$' or $`b$' and hence, all strings starting with $`b$' will reach the state $q_1$ and are rejected. Strings starting with $`a$' will go to the state $`q_2$' and since $\Sigma_{q_2}=\{a\}$, at the state $q_2$ the automaton can jump over a sequence of $`b$'s and will delete an $`a$' and go to the state $q_3$. Similarly, since $\Sigma_{q_3}=\{b\}$, at state $q_3$ the automaton can jump over a sequence of $`a$'s and will delete $`b$' and go to state $q_2$. For example, consider a string `$abbaa$'. The sequence of transitions is 
$$q_0abbaa \circlearrowright q_2bbaa \circlearrowright q_3abb  \circlearrowright q_2ba \circlearrowright q_3b \circlearrowright q_2.$$

Hence, the language accepted by the automaton is 

$$L_R(\mathcal{A})=\{aw~:~ |w|_a=|w|_b, w \in \{a,b\}^*\}.
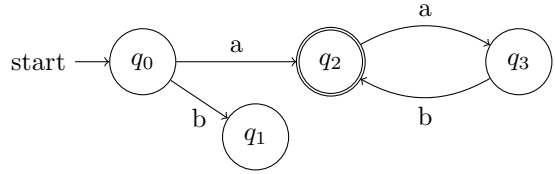
$$
$$
\begin{tikzpicture}
\node[state, initial] (q0) {$q_0$};
\node[state, accepting, right of=q0, xshift=1.5cm] (q2) {$q_2$};
\node[state, below of=q0, xshift=1.5cm] (q1) {$q_1$};
\node[state, right of=q2, xshift=1.5cm] (q3) {$q_3$};
\path[->]
(q0) edge[above] node{a} (q2)
(q0) edge[below] node{b} (q1)
(q2) edge[bend left, above] node{a} (q3)
(q3) edge[bend left, below] node{b} (q2);
\end{tikzpicture}
$$
\end{example}

\section{Generalized Linear One-Way Jumping Finite Automata}\label{glowjfa}
In a right one-way jumping finite automaton, the read head moves in one direction only and starts from the leftmost symbol of the input word. It moves from left to right (and possibly jumps over parts of the input) and upon reaching the end of the input word the automaton will start reading the remaining concatenated input freshly from the last visited state. The computation continues until all the letters are read or the automaton is stuck in a state in which it cannot delete any letter of the remaining input. If a transition is defined for the current state and the next letter to be read, then the automaton deletes the symbol. If not, but in the remaining input there are letters for which a transition is defined from the current state, the read head jumps to the nearest such letter to the right for its reading.

We extend this definition based on reading length of a word and define a generalized right/left linear one-way jumping finite automaton. A generalized right linear one-way jumping finite automaton deletes an input word from left to right. The automaton starts deleting the input word with the reading head at the leftmost position of the input word. The automaton can jump over a part of the input word. It reads the nearest available subword of the word present on the input tape. If there is no transition available for a state, the automaton returns(head of the automaton returns) to the leftmost position of the current input and continues its computation. The formal definitions of generalized right linear one-way jumping finite automaton and generalized left linear one-way jumping finite automaton are given below.

\begin{definition}\label{grld}
A generalized right linear one-way jumping finite automaton$~~$ ($\it{GRLOWJFA}$) is a tuple $\mathcal{A}=(\Sigma,Q,q_0,F, R)$, where $\Sigma,Q,q_0,F$ are same as $\it{ROWJFA}$ and $R \subset Q \times \Sigma^+ \times Q$ is a finite set of rules, where for a state $p \in Q$ and a word $w \in \Sigma^+$, there is at most one $q \in Q$ such that $(p,w,q) \in R$.  By a rule $(p,w,q) \in R$, we mean that the automaton goes to the state $q$ from the state $p$ after deleting the word `$w$'. The automaton is deterministic in the sense that for a state $p$ and for a word $w \in \Sigma^+$ we have at most one $q \in Q$ such that $(p,w,q) \in R$. A configuration of the automaton $\mathcal{A}$ is a string of $\Sigma^* Q\Sigma^*$. The generalized right linear one-way jumping relation, denoted as $\curvearrowright_{\mathcal{A}}$, or just $\curvearrowright$ if it is clear which $\it{GRLOWJFA}$ is being referred, over $\Sigma^*Q\Sigma^*$ is defined as follows: For a state $p \in Q$, set $\Sigma_p=\Sigma_{R,p}=\{w \in \Sigma^+ 
: (p,w,q) \in R$ for some $q \in Q\}$.

\begin{itemize}
    \item[1.] Let $t,u,v \in \Sigma^*$ and $(p,x,q) \in R$. Then the $\it{GRLOWJFA}$ $\mathcal{A}$ makes a jump from the configuration $tpuxv$ to the configuration $tuqv$, written as $$tpuxv \curvearrowright tuqv$$ 
if $u$ does not contain any word from $\Sigma_p$ as a subword, i.e., $u \neq u'wu''$,
where $u',u'' \in \Sigma^*,~w \in \Sigma_p$ and $u_2x_1 \neq x$, where $u_2, x_1 \in \Sigma^+$ and $u=u_1u_2,~x=x_1x_2$ for some $u_1,x_2 \in \Sigma^*$.
\item[2.] Let $x \in \Sigma^+,y \in \Sigma^*$ and $y$ does not contain any word from $\Sigma_p$ as a subword, i.e., $y \neq y_1wy_2$, where $y_1,y_2 \in \Sigma^*$ and $w \in \Sigma_p$, then the $\it{GRLOWJFA}$ $\mathcal{A}$ makes a jump from the configuration $xpy$ to the configuration $pxy$, written as $$xpy \curvearrowright pxy.$$ 
\end{itemize} 

 The language accepted by the $\it{GRLOWJFA}$ $\mathcal{A}$ is $$L_{GRL}(\mathcal{A})=\{w \in \Sigma^* : q_0w \curvearrowright^* q_f ~ for~some ~ q_f \in F\}.$$
\end{definition}
Similar to that of {\it GRLOWJFA}, we define the notion of a generalized left linear one-way jumping finite automaton as below.

\begin{definition}\label{glld}
A generalized left linear one-way jumping finite automaton denoted by $\it{GLLOWJFA}$ is a tuple $\mathcal{A}=(\Sigma,Q,q_0,F, R)$, where $\Sigma,Q,q_0,F$ are same as $\it{ROWJFA}$ and $R \subset Q \times \Sigma^+ \times Q$ is a finite set of rules, where for a state $p \in Q$ and a word $w \in \Sigma^+$, there is at most one $q \in Q$ such that $(q,w,p) \in R$.  By a rule $(q,w,p) \in R$, we mean that the automaton goes to the state $q$ from the state $p$ after deleting the word `$w$'. The automaton is deterministic in the sense that for a state $p$ and for a word $w \in \Sigma^+$ we have at most one $q \in Q$ such that $(q,w,p) \in R$. A configuration of the automaton $\mathcal{A}$ is a string of $\Sigma^*Q\Sigma^*$. The generalized left linear one-way jumping relation, denoted as $\prescript{}{{\mathcal{A}}}{\curvearrowleft}$, or just $\curvearrowleft$ if it is clear which $\it{GLLOWJFA}$ is being referred, over $\Sigma^*Q\Sigma^*$ is defined as follows: For a state $p \in Q$, set $\Sigma_p=\Sigma_{R,p}=\{w \in \Sigma^+ 
: (q,w,p) \in R$ for some $q \in Q\}$.
\begin{itemize}
    \item[1.] Let $t,u,v \in \Sigma^* $ and $(q,x,p) \in R$. Then, the $\it{GLLOWJFA}$ $\mathcal{A}$ makes a jump from the configuration $vxupt$ to the configuration $vqut$, written as $$vqut \curvearrowleft vxupt$$ 
if $u$ does not contain any word from $\Sigma_p$ as a subword, i.e., $u \neq u'wu''$,
where $u',u'' \in \Sigma^*,~w \in \Sigma_p$ and $x_2u_1 \neq x$, where $u_1, x_2 \in \Sigma^+$ and $u=u_1u_2,~x=x_1x_2$ for some $u_2,x_1 \in \Sigma^*$. 

\item[2.] Let $x \in \Sigma^+,y \in \Sigma^*$ and $y$ does not contain any word from $\Sigma_p$ as a subword, i.e., $y \neq y_1wy_2$, where $y_1,y_2 \in \Sigma^*$ and $w \in \Sigma_p$, then the $\it{GLLOWJFA}$ $\mathcal{A}$ makes a jump from the configuration $ypx$ to the configuration $yxp$, written as $$yxp \curvearrowleft ypx.$$
\end{itemize}
The language accepted by the $\it{GLLOWJFA}$ $\mathcal{A}$ is $$L_{GLL}(\mathcal{A})=\{w \in \Sigma^* : q_f~ \prescript{*}{}{\curvearrowleft}~ wq_0 ~ for~some ~ q_f \in F\}.$$

\end{definition}

We illustrate Definition \ref{grld} with the following example.

\begin{example}\label{exrl}
Consider the following automaton $\mathcal{A}=(\{a,b\},\{q_0,q_1\},q_0,~~$  $\{q_1\},R)$, where $R$ is depicted in the figure below.

$$
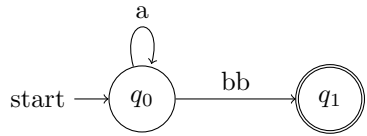
\begin{tikzpicture}
\node[state, initial] (q0) {$q_0$};
\node[state, accepting, right of=q0, xshift=1.5cm] (q1) {$q_1$};
\path[->] (q0) edge[loop above] node{a} (q0)
(q0) edge  [above] node{bb} (q1);
\end{tikzpicture}
$$
Here, $\Sigma_{q_0}=\{a,bb\}$ and $\Sigma_{q_1}= \emptyset$.
We first consider the automaton as a $\it{GRLOWJFA}$. The automaton can read arbitrary number of $a$'s at state $q_0$ and can jump only one $b$ at $q_0$. Note that as $\Sigma_{q_0}=\{a,bb\}$, the automaton at $q_0$ can neither jump over an $a$ or $bb$. Once it reads $bb$, it reaches the final state $q_1$ and no more transition is possible. 
Consider the word $a^lba^mba^n$, where $l,n \geq 0, m \geq 1$. Then,

 $$q_0a^lba^mba^n \curvearrowright^* q_0ba^mba^n \curvearrowright bq_0a^{m-1}ba^n $$
 $$\curvearrowright^*
 bq_0ba^n \curvearrowright bbq_0a^{n-1} \curvearrowright^* bbq_0 \curvearrowright q_0bb \curvearrowright q_1$$
and,
$$L_{GRL}(\mathcal{A})=\{a^nbb~|~ n \geq 0 \} \cup \{a^lba^mba^n~|~ l,n \geq 0, m \geq 1\}.$$
Similarly, the language accepted by the automaton when it is considered as $\it{GLLOWJFA}$ is  
 $$L_{GLL}(\mathcal{A})=\{bba^n~|~ n \geq 0 \} \cup \{a^lba^mba^n~|~ l,n \geq 0, m \geq 1\}.$$

\end{example}

Now, we give an example which shows that the intersection of the language classes $\bf{GRLOWJ}$ and $\bf{GLLOWJ}$ is non empty, where $\bf{GRLOWJ}$ and $\bf{GLLOWJ}$ represent the language classes accepted by $\it{GRLOWJFA}$ and $\it{GLLOWJFA}$, respectively.

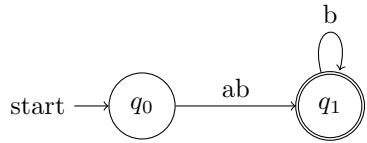
\begin{example}

$$
\begin{tikzpicture}
\node[state, initial] (q0) {$q_0$};
\node[state, accepting, right of=q0, xshift=1.5cm] (q1) {$q_1$};
\path[->] (q0) edge[above] node{ab} (q1)
(q1) edge[loop above] node{b} (q1);
\end{tikzpicture}
$$
When considered as $\it{GRLOWJFA}$, the automaton will jump to $ab$ from the configuration $q_0b^nabb^m$, it will delete $ab$ and will go to the configuration $b^nq_1b^m$. Then, it will go to the configuration $b^nq_1$ using repeated application of the rule $(q_1,b,q_1)$. The automaton will jump to the configuration $q_1b^n$ from the configuration $b^nq_1$ and using the repeated application of the rule $(q_1,b,q_1)$ the automation will reach $q_1$ and hence, accepts the word $b^nab^m$. Consider a word $b^mabb^n$, where $m,n \geq 0$. Then, 
$$q_0b^mabb^n \curvearrowright b^mq_1b^n \curvearrowright^* b^mq_1 \curvearrowright q_1b^m \curvearrowright^* q_1.$$

Similarly, the case of $\it{GLLOWJFA}$ can be explained. Note that the set of rules is $R=\{(q_1,ab,q_0),(q_1,b,q_1)\}$, when the automaton is considered as $\it{GLLOWJFA}$. Hence,
    $$L_{GRL}(\mathcal{A})=L_{GLL}(\mathcal{A})=\{b^mabb^n~|~ n,m \geq 0\}.$$

\end{example}

\begin{note}
$\bf{GRLOWJ \cap GLLOWJ} \neq \emptyset$.
\end{note}

\section{\bf{GLOWJ} and \bf{OWJ}}\label{glowjowj}

In this section, we compare the language classes $\bf{ROWJ}$ and $\bf{GRLOWJ}$ as well as the language classes $\bf{LOWJ}$ and $\bf{GLLOWJ}$, here $\bf{ROWJ}$ and $\bf{LOWJ}$ represent the language classes accepted by $\it{ROWJFA}$ and $\it{LOWJFA}$, respectively. We show that the language class $\bf{ROWJ}$ is a proper subset of the language class $\bf{GRLOWJ}$ and the language class $\bf{LOWJ}$ is a proper subset of the language class $\bf{GLLOWJ}$.

By definitions of $\it{GRLOWJFA}$ and $\it{ROWJFA}$, it is clear that when the rules of a $\it{GRLOWJFA}$ satisfy the condition: if $(p,w,q) \in R$, then $|w|=1$, then the $\it{GRLOWJFA}$ is same as $\it{ROWJFA}$. Hence, we have $\bf{ROWJ}$ $\subseteq$ $\bf{GRLOWJ}$. Similarly, $\bf{LOWJ}$ $\subseteq$ $\bf{GLLOWJ}$. Now, we give a language which is accepted by the model $\it{GRLOWJFA}$ but not by the model $\it{ROWJFA}$.
\begin{example}\label{nonrowj}
Consider the generalized right linear one-way jumping finite automaton $\mathcal{A}=(\{a,b\}, \{q_0,q_1,q_2,q_3,q_4\},$ $q_0,\{q_0,q_1,q_2,q_4\}, R)$, where the set of rules $R=\{(q_0,a,q_1),(q_1,b,q_2),(q_2,a,q_3),$ $(q_3,b,q_2),(q_0,aa,q_4),$ $(q_4,a,q_4)\}$. 
The language accepted by the automaton is $L_{GRL}(\mathcal{A})=\{w \in \{a,b\}^*~|~|w|_a=|w|_b$ or $|w|_b=0\}$.
\end{example}
It was proved in \cite{porowjfa} that the language of Example \ref{nonrowj} cannot be accepted by any $\it{ROWJFA}$. Hence, we have the following result.

\begin{lemma}\label{rowgrlow}
There exists a language which is accepted by a $\it{GRLOWJFA}$ but not by any $\it{ROWJFA}$ and hence, $\bf{ROWJ \subset GRLOWJ}$.
\end{lemma}

Now, we give an example of a $\it{GLLOWJFA}$ and we show that the language of the automaton cannot be accepted by any $\it{LOWJFA}$.

\begin{example}\label{dyckgll}
Consider a $\it{GLLOWJFA}$ $\mathcal{A}=(\{a,b\}, \{q_0\}, q_0, \{q_0\},\{(q_0,ab,q_0)\})$. One can verify that the language accepted by the automaton is the Dyck language, $D$. 
\end{example}
We show the Dyck language cannot be accepted by any $\it{LOWJFA}$.

\begin{lemma}
The Dyck language cannot be accepted by any $\it{LOWJFA}$.
\end{lemma}
\begin{proof}
Let the Dyck language, $D$, accepted by a $\it{LOWJFA}$, say $\mathcal{A}=(\Sigma,$ $Q,q_0,F,R)$. Then $L(\mathcal{A})=D$. Take a positive integer $k$, where $k > |Q|$. Since $a^nb^n \in D$ for all $n \geq 0$, therefore $a^kb^k \in D$ and hence, $a^kb^k \in L(\mathcal{A})$. Then there exists a sequence of transitions such that $ q_f \prescript{*}{}{\circlearrowleft}~ a^kb^kq_0$, where $q_f \in F$. 

First we show the automaton cannot delete any `$a$' without deleting a $`b$' from the configuration $a^kb^kq_0$. Suppose the automaton deletes an $`a$' from the configuration $a^kb^kq_0$, then there exist a state $q' \in Q$ and $(q',a,q_0) \in R$, $b \not\in \Sigma_{q_0}$ such that $q_f \prescript{*}{}{\circlearrowleft}~ b^ka^{k-1}q' \circlearrowleft a^kb^kq_0$. Then, the word $b^ka^k \in L(\mathcal{A})$ because $q_f \prescript{*}{}{\circlearrowleft}~ b^ka^{k-1}q' \circlearrowleft b^ka^kq_0$. But $b^ka^k \not\in D$.

Now, we show that the automaton will have to delete all $b$'s before deleting any $`a$' starting from the configuration $a^kb^kq_0$. Suppose the automaton deletes $t$ $b$'s starting from the configuration $a^kb^kq_0$ before deleting an `$a$', where $1 \leq t < k$. Then there exists a sequence of transitions such that $q_f \prescript{*}{}{\circlearrowleft}~ b^{k-t}a^{k-1}q'' \circlearrowleft a^kb^{k-t}q' \prescript{*}{}{\circlearrowleft}~ a^kb^kq_0$, where $q',q'' \in Q$, $(q'',a,q') \in R$ and $b \not\in \Sigma_{q'}$. Then, the word $b^{k-t}a^kb^t$ will also be in $L(\mathcal{A})$ because $q_f \prescript{*}{}{\circlearrowleft}~ b^{k-t}a^{k-1}q'' \circlearrowleft b^{k-t}a^kq' \prescript{*}{}{\circlearrowleft}~ b^{k-t}a^kb^tq_0$. But $b^{k-t}a^kb^t \not\in D$.

Hence, the automaton will have to delete all $b$'s starting from the configuration $a^kb^kq_0$ before deleting any `$a$'. But in that case the automaton will have to loop because $|a^kb^k|_b > |Q|$ and hence, the words of the form $a^kb^{k+l}$ will be in $L(\mathcal{A})$, where $l \geq 0$. But $a^kb^{k+l} \not\in D$ for all $l \geq 0$.

Hence, the Dyck language cannot be accepted by any $\it{LOWJFA}.$ 
\end{proof}

From the above Lemma we have the following result.

\begin{lemma}
There exists a language which is accepted by a $\it{GLLOWJFA}$ but not by any $\it{LOWJFA}$ and hence, $\bf{LOWJ \subset GLLOWJ}$.
\end{lemma}

\section{$\it{GRLOWJFA}$ and $\it{GLLOWJFA}$}\label{rightleft}

In this section, we compare the language classes $\bf{GRLOWJ}$ and $\bf{GLLOWJ}$. We also establish a relationship between the languages of the class $\bf{GRLOWJ}$ and the languages of the class $\bf{GLLOWJ}$.

First we show that for every language $L_1 \in \bf{GRLOWJ}$ $(\bf{GLLOWJ})$, there exists a language $L_2 \in \bf{GLLOWJ}$ $(\bf{GRLOWJ}$ resp) such that $L_1^R=L_2$.

\begin{proposition}
For a given language $L_1 \in \bf{GRLOWJ}$, there exists a language $L_2 \in \bf{GLLOWJ}$ such that $L_1^R=L_2$ and vice versa.
\end{proposition}
\begin{proof}
Let $L_1 \in \bf{GRLOWJ}$. Then there exists a $\it{GRLOWJFA}$, say $\mathcal{A}_1=(\Sigma,Q,q_0,F,R)$, such that $L_1=L(\mathcal{A}_1)$. We construct a $\it{GLLOWJFA}$, say $\mathcal{A}_2$, as: $\mathcal{A}_2=(\Sigma,Q,q_0,F,R')$, where $R'=\{(q,w^R,p)~|~ (p,w,q) \in R\}$. Let $L(\mathcal{A}_2)=L_2$. Claim, $L_1^R=L_2$. Under this construction we have the following lemmas:

\begin{lemma}\label{xpyux}
If $x,y,z \in \Sigma^*$, $u \in \Sigma^+$ and $p,q \in Q$, then $xpyuz \curvearrowright_{\mathcal{A}_1} xyqz$ if and only if $z^Rqy^Rx^R$ $\prescript{}{{\mathcal{A}_2}}{\curvearrowleft}$  $z^Ru^Ry^Rpx^R$.
\end{lemma}
\begin{proof}
Let $x,y,z \in \Sigma^*$, $u \in \Sigma^+$ and $p,q \in Q$. Now, $xpyuz \curvearrowright_{\mathcal{A}_1} xyqz$ if and only if $(p,u,q) \in R$, $y$ does not contain any word from $\Sigma_{R,p}$ as a subword and $y''u' \neq u$, where $y''$ is a nonempty suffix of $y$ and $u'$ is a nonempty prefix of $u$ if and only if $(q,u^R,p) \in R'$, $y^R$ does not contain any word from $\Sigma_{R',p}$ as a subword and $u^{(2)}y^{(1)} \neq u^R$, where $y^{(1)}$ is a nonempty prefix of $y^R$ and $u^{(2)}$ is a nonempty suffix of $u^R$ if and only if $z^Rqy^Rx^R$ $\prescript{}{{\mathcal{A}_2}}{\curvearrowleft}$ $z^Ru^Ry^Rpx^R$. Hence, for $x,y,z \in \Sigma^*$, $u \in \Sigma^+$ and $p,q \in Q$, $xpyuz \curvearrowright_{\mathcal{A}_1} xyqz$ if and only if $z^Rqy^Rx^R$ $\prescript{}{{\mathcal{A}_2}}{\curvearrowleft}$   $z^Ru^Ry^Rpx^R$. 
\end{proof}

\begin{lemma}\label{xpy}
If $x \in \Sigma^+$, $y \in \Sigma^*$ and $p \in Q$, then $xpy \curvearrowright_{\mathcal{A}_1} pxy$ if and only if $y^Rx^Rp$ $\prescript{}{{\mathcal{A}_2}}{\curvearrowleft}$ $y^Rpx^R$.
\end{lemma}
\begin{proof}
Let $x \in \Sigma^+$, $y \in \Sigma^*$ and $p \in Q$. Now, $xpy \curvearrowright_{\mathcal{A}_1} pxy$ if and only if $y$ does not contain any word from $\Sigma_{R,p}$ as a subword if and only if $y^R$ does not contain any word $\Sigma_{R',p}$ as a subword if and only if $y^Rx^Rp$ $\prescript{}{{\mathcal{A}_2}}{\curvearrowleft}$ $y^Rpx^R$. Hence, for $x \in \Sigma^+$, $y \in \Sigma^*$ and $p \in Q$, $xpy \curvearrowright_{\mathcal{A}_1} pxy$ if and only if $y^Rx^Rp$ $\prescript{}{{\mathcal{A}_2}}{\curvearrowleft}$ $y^Rpx^R$.
\end{proof}

Now, from Lemmas \ref{xpyux} and \ref{xpy}, $w \in L_1^R$ if and only if $q_0w^R \curvearrowright_{\mathcal{A}_1}^* q_f$, where $q_f \in F$, if and only if $q_f$ $\prescript{*}{{\mathcal{A}_2}}{\curvearrowleft}$ $wq_0$ if and only if $w \in L_2$. Hence, $L_1^R=L_2$.

\end{proof}

\begin{corollary}
If a language $L \in \bf{GRLOWJ \cap GLLOWJ}$, then the reversal of the language $L^R$ $\in$ $\bf{GRLOWJ \cap GLLOWJ}$.
\end{corollary}

Now, we compare the language classes $\bf{GRLOWJ}$ and $\bf{GLLOWJ}$. We give a language which is in the class $\bf{GLLOWJ}$ and prove that the language is not in the class $\bf{GRLOWJ}$ which proves $\bf{GLLOWJ}$ $\not\subseteq$ $\bf{GRLOWJ}$. Similarly, it can be proved $\bf{GRLOWJ}$ $\not\subseteq$ $\bf{GLLOWJ}$. Hence, we conclude the language classes $\bf{GRLOWJ}$ and $\bf{GLLOWJ}$ are incomparable.

\begin{example}\label{Dc}
Consider the $\it{GLLOWJFA}$ $\mathcal{A}=(\{a,b,c\},\{q_0,q_1,q_2\},q_0,\{q_1\},R)$, where $R=\{(q_1,c,q_0),(q_1,ab,q_1),(q_2,a,q_0),(q_2,b,q_0)\}$. The language accepted by the automaton is $L_{GLL}(\mathcal{A})=Dc$, where $D$ is the Dyck language.
\end{example}

Now, we prove that the language $Dc$ is not in the class $\bf{GRLOWJ}$.

\begin{lemma}\label{dyckcon}
There does not exist any $\it{GRLOWJFA}$ that accepts the language $Dc$, where $D$ is the Dyck language.
\end{lemma}
\begin{proof}
If $Dc \in \bf{GRLOWJ}$, then there exists a $\it{GRLOWJFA}$, say $\mathcal{A}=(\Sigma,Q,q_0,$ $F,R)$, such that $L(\mathcal{A})=Dc$. Let $l=|Q|$ and $m=max\{|w|~|~ (p,w,q)$ $\in R\}$. Choose a natural number $n > 5lm$. Clearly, $a^nb^nc \in Dc=L(\mathcal{A})$. Hence, there exists a sequence of transitions such that $q_0a^nb^nc \curvearrowright^* q_f$, where $q_f \in F$.

From the configuration $q_0a^nb^nc$, the automaton can go to one of the following configurations: $qa^{n-i}b^nc$, $a^{n-i}qb^{n-j}c$, $a^nqb^{n-j}c$, $a^nb^{n-j}q$, $a^nb^nq$, where $i,j \geq 1$, $q \in Q$. 
\vspace{0.2cm}

The `$c$' cannot be deleted by the automaton before deleting all $a$'s and $b$'s. If this happens, then there would be a sequence of transitions such that $q_0a^nb^nc \curvearrowright^* a^{n-k_1}b^{n-k_2}q \curvearrowright qa^{n-k_1}b^{n-k_2} \curvearrowright^* q_f$, where $k_1,k_2 \geq 0$ and at least one of them not equal to $n$, $q \in Q$. But then, the following sequence is also possible: $q_0a^{i_1}b^{j_1}a^{i_2}b^{j_2} \cdots a^{i_t}b^{j_t}ca^{n-k_1}b^{n-k_2} \curvearrowright^* qa^{n-k_1}b^{n-k_2} \curvearrowright^* q_f$, where $\sum_{k=1}^t i_k =k_1, \sum_{k=1}^t j_k=k_2$. Hence, the word $a^{i_1}b^{j_1}a^{i_2}b^{j_2} \cdots a^{i_t}b^{j_t}ca^{n-k_1}b^{n-k_2}$ $\in L(\mathcal{A})$. Since at least one of $(n-k_1)$ and $(n-k_2)$ is non-zero, therefore\\ $ a^{i_1}b^{j_1}a^{i_2}b^{j_2} \cdots a^{i_t}b^{j_t}ca^{n-k_1}b^{n-k_2} \not\in Dc$. Hence, the `$c$' cannot be deleted by the the automaton before deleting all $a$'s and $b$'s. 

Therefore, the automaton cannot go to the configuration $a^nb^{n-j}q$ or $a^nb^nq$ from the configuration $q_0a^nb^nc$. 
\vspace{0.2cm}

If the automaton goes to the configuration $a^nqb^{n-j}c$. The automaton cannot delete all $b$'s starting from the configuration $a^nqb^{n-j}c$ because in order to delete all $b$'s it will have to loop and in that case mismatch in the indices of $`a$' and $`b$' can be created.  Therefore, the automaton will use the following sequence of transitions to accept the word $a^nb^nc$: $q_0a^nb^nc$ $\curvearrowright$ $a^nqb^{n-j}c$  $\curvearrowright^*$ $a^nq'b^{n-j-j'}c$ $\curvearrowright$ $q'a^nb^{n-j-j'}c$ $\curvearrowright^*$ $q_f$, where $j' \geq 0$, $q' \in Q$ and $b^t \not\in \Sigma_{q'}$ for $1 \leq t \leq n-j-j'$, $b^tc \not\in \Sigma_{q'}$ for $0 \leq t \leq n-j-j'$. But, then the following sequence of transitions is also possible: $q_0b^{j+j'}a^nb^{n-j-j'}c$ $\curvearrowright$ $qb^{j'}a^nb^{n-j-j'}c$ $\curvearrowright^*$ $q'a^nb^{n-j-j'}c$ $\curvearrowright^*$ $q_f$. Hence, the word $b^{j+j'}a^nb^{n-j-j'}c$ $\in L(\mathcal{A)}$ but $b^{j+j'}a^nb^{n-j-j'}c$ $\not\in Dc$. Hence, the automaton cannot go to the configuration $a^nqb^{n-j}c$ from the configuration $q_0a^nb^nc$. 
\vspace{0.2cm}

If the automaton goes to the configuration $a^{n-i}qb^{n-j}c$ from the configuration $q_0a^nb^nc$, then $q_0a^nb^nc$ $\curvearrowright$ $a^{n-i}qb^{n-j}c$ $\curvearrowright^*$ $a^{n-i}q^{(1)}b^{n-j-j_1}c$, where $q^{(1)} \in Q,j_1 \geq 0$. Since `$c$' cannot be deleted before deleting all $a$'s and $b$'s, therefore
$q_0a^nb^nc$ $\curvearrowright$ $a^{n-i}qb^{n-j}c$ $\curvearrowright^*$ $a^{n-i}q^{(1)}b^{n-j-j_1}c$ $\curvearrowright$ $q^{(1)}a^{n-i}b^{n-j-j_1}c$, where $b^t \not\in \Sigma_{q^{(1)}}$ for $1 \leq t \leq n-j-j_1$ and $b^tc \not\in \Sigma_{q^{(1)}}$ for $0 \leq t \leq n-j-j_1$. From the configuration $q^{(1)}a^{n-i}b^{n-j-j_1}c$ the automaton can go to the configuration $q^{(2)}a^{n-i-i_1}b^{n-j-j_1}c$, where $q^{(2)} \in Q,i_1 \geq 1$ or $a^{n-i-i_1}q^{(2)}b^{n-j-j_1-j_2}c$, where $q^{(2)} \in Q$, $i_1,j_2 \geq 1$, $a^t \not\in \Sigma_{q^{(1)}}$ for $1 \leq t \leq n-i-i_1$. Hence, we have the following two sequence of transitions:

$1.$ $q_0a^nb^nc$ $\curvearrowright$ $a^{n-i}qb^{n-j}c$ $\curvearrowright^*$ $a^{n-i}q^{(1)}b^{n-j-j_1}c$ $\curvearrowright$ $q^{(1)}a^{n-i}b^{n-j-j_1}c$ $\curvearrowright$\\ $q^{(2)}a^{n-i-i_1}b^{n-j-j_1}c$ $\curvearrowright^*$ $q_f$. But, then the following sequence is also possible: $q_0a^{n-i_1}b^nca^{i_1}$ $\curvearrowright$ $a^{n-i_1-i}qb^{n-j}ca^{i_1}$ $\curvearrowright^*$ $a^{n-i_1-i}q^{(1)}b^{n-j-j_1}ca^{i_1}$ $\curvearrowright$ $a^{n-i_1-i}b^{n-j-j_1}cq^{(2)}$ $\curvearrowright$ $q^{(2)}a^{n-i-i_1}b^{n-j-j_1}c$ $\curvearrowright^*$ $q_f$. Hence, the word $a^{n-i_1}b^nca^{i_1}$ $\in$ $L(\mathcal{A})$ but $a^{n-i_1}b^nca^{i_1}$ $\not\in$ $Dc$.

$2.$ $q_0a^nb^nc$ $\curvearrowright$ $a^{n-i}qb^{n-j}c$ $\curvearrowright^*$ $a^{n-i}q^{(1)}b^{n-j-j_1}c$ $\curvearrowright$ $q^{(1)}a^{n-i}b^{n-j-j_1}c$ $\curvearrowright$\\ $a^{n-i-i_1}q^{(2)}b^{n-j-j_1-j_2}c$ $\curvearrowright^*$ $a^{n-i-i_1}q^{(3)}b^{n-j-j_1-j_2-j_3}c$ $\curvearrowright$ $q^{(3)}a^{n-i-i_1}b^{n-j-j_1-j_2-j_3}c$, where $j_3 \geq 0, q^{(3)} \in Q$, $b^t \not\in \Sigma_{q^{(3)}}$ for $1 \leq t \leq n-j-j_1-j_2-j_3$ and $b^tc \not\in \Sigma_{q^{(3)}}$ for $0 \leq t \leq n-j-j_1-j_2-j_3$. If $q^{(3)}$ deletes only $a$'s, then this case becomes similar to case $1.$ and hence, there exists a word $w \in L(\mathcal{A})$ but $w \not\in Dc$. Therefore, $q^{(3)}$ will have to delete $a^{i_2}b^{j_4}$, where $i_2,j_4 \geq 1$ and hence, $q_0a^nb^nc$ $\curvearrowright$ $a^{n-i}qb^{n-j}c$ $\curvearrowright^*$ $a^{n-i}q^{(1)}b^{n-j-j_1}c$ $\curvearrowright$ $q^{(1)}a^{n-i}b^{n-j-j_1}c$ $\curvearrowright$\\ $a^{n-i-i_1}q^{(2)}b^{n-j-j_1-j_2}c$ $\curvearrowright^*$ $a^{n-i-i_1}q^{(3)}b^{n-j-j_1-j_2-j_3}c$ $\curvearrowright$ $q^{(3)}a^{n-i-i_1}b^{n-j-j_1-j_2-j_3}c$ $\curvearrowright$ $a^{n-i-i_1-i_2}q^{(4)}b^{n-j-j_1-j_2-j_3-j_4}c$ $\curvearrowright^*$ $a^{n-i-i_1-i_2}q^{(5)}b^{n-j-j_1-j_2-j_3-j_4-j_5}c$ $\curvearrowright$\\ $q^{(5)}a^{n-i-i_1-i_2}b^{n-j-j_1-j_2-j_3-j_4-j_5}c$ $\curvearrowright^*$ $q_f$, where $j_5 \geq 0, q^{(4)},q^{(5)} \in Q$, $b^t \not\in \Sigma_{q^{(5)}}$ for $1 \leq t \leq n-j-j_1-j_2-j_3-j_4-j_5$ and $b^tc \not\in \Sigma_{q^{(5)}}$ for $0 \leq t \leq n-j-j_1-j_2-j_3-j_4-j_5$. But, then the following sequence of transition is also possible: \\$q_0a^{n-i_1-i_2}b^{n-j_2-j_3-j_4-j_5}a^{i_1}b^{j_2+j_3}ca^{i_2}b^{j_4+j_5}$ $\curvearrowright$\\ $a^{n-i_1-i_2-i}qb^{n-j_2-j_3-j_4-j_5-j}a^{i_1}b^{j_2+j_3}ca^{i_2}b^{j_4+j_5}$ $\curvearrowright^*$\\ $a^{n-i_1-i_2-i}q^{(1)}b^{n-j_2-j_3-j_4-j_5-j-j_1}a^{i_1}b^{j_2+j_3}ca^{i_2}b^{j_4+j_5}$ $\curvearrowright$\\ $a^{n-i_1-i_2-i}b^{n-j_2-j_3-j_4-j_5-j-j_1}q^{(2)}b^{j_3}ca^{i_2}b^{j_4+j_5}$ $\curvearrowright^*$ \\ $a^{n-i_1-i_2-i}b^{n-j_2-j_3-j_4-j_5-j-j_1}q^{(3)}ca^{i_2}b^{j_4+j_5}$ $\curvearrowright$ \\
$a^{n-i_1-i_2-i}b^{n-j_2-j_3-j_4-j_5-j-j_1}cq^{(4)}b^{j_5}$ $\curvearrowright^*$\\ $a^{n-i_1-i_2-i}b^{n-j_2-j_3-j_4-j_5-j-j_1}cq^{(5)}$ $\curvearrowright$\\ $q^{(5)}a^{n-i_1-i_2-i}b^{n-j_2-j_3-j_4-j_5-j-j_1}c$ $\curvearrowright^*$ $q_f$.\\ Hence, the word $w=a^{n-i_1-i_2}b^{n-j_2-j_3-j_4-j_5}a^{i_1}b^{j_2+j_3}ca^{i_2}b^{j_4+j_5}$ $\in$ $L(\mathcal{A})$ but $w \not\in Dc$.

\vspace{0.2cm}
If the automaton goes to the configuration $qa^{n-i}b^nc$ from the configuration $q_0a^nb^nc$. The automaton cannot delete all $a$'s from the configuration $qa^{n-i}b^nc$ otherwise the automation will have to loop and in this case mismatch in the indices of `$a$' and `$b$' can be created. Hence, the automaton will make the following sequence of transitions: $q_0a^nb^nc$ $\curvearrowright$ $qa^{n-i}b^nc$ $\curvearrowright^*$ $q^{(1)}a^{n-i-i_1}b^nc$ $\curvearrowright$\\ $a^{n-i-i_1-i_2}q^{(2)}b^{n-j_1}c$ $\curvearrowright^*$ $a^{n-i-i_1-i_2}q^{(3)}b^{n-j_1-j_2}c$ $\curvearrowright$ $q^{(3)}a^{n-i-i_1-i_2}b^{n-j_1-j_2}c$, where $i_1,i_2,j_2 \geq 0$, $j_1 \geq 1$, $b^t \not\in \Sigma_{q^{(3)}}$ for $1 \leq t \leq n-j_1-j_2$ and $b^tc \not\in \Sigma_{q^{(3)}}$ for $0 \leq t \leq n-j_1-j_2$. From the configuration $q^{(3)}a^{n-i-i_1-i_2}b^{n-j_1-j_2}c$ the automaton can go to the configuration $q^{(4)}a^{n-i-i_1-i_2-i_3}b^{n-j_1-j_2}c$, where $q^{(4)} \in Q,i_3 \geq 1$ or $a^{n-i-i_1-i_2-i_3}q^{(4)}b^{n-j_1-j_2-j_3}c$, where $q^{(4)} \in Q$, $i_3,j_3 \geq 1$, $a^t \not\in \Sigma_{q^{(3)}}$ for $1 \leq t \leq n-i-i_1-i_2-i_3$. These cases are similar to the above cases 1. and 2. Hence, there will be word $w \in \Sigma^*$ such that $w$ $\in$ $L(\mathcal{A})$ but $w \not\in$ $Dc$.

Hence, the language $Dc \not\in \bf{GRLOWJ}$.

\end{proof}

From Example \ref{Dc} and Lemma \ref{dyckcon}, we conclude $\bf{GLLOWJ \not\subseteq GRLOWJ}$. Similarly, it can be shown that $\bf{GRLOWJ \not\subseteq GLLOWJ}$. Hence, we have the following result.

\begin{proposition}
The language classes $\bf{GRLOWJ}$ and $\bf{GLLOWJ}$ are incomparable.
\end{proposition}

\section{GLOWJFA and Chomsky hierarchy}\label{chomsky}

In this section, we compare the language classes $\bf{GRLOWJ}$ with the language classes of Chomsky hierarchy. $\bf{REG}$, $\bf{CF}$ and $\bf{CS}$ represent the class of regular languages, context free languages and context sensitive languages, respectively. We show that the language class $\bf{REG}$ is a proper subset of the class $\bf{GRLOWJ}$, the language classes $\bf{CF}$ and $\bf{GRLOWJ}$ are incomparable and the class $\bf{GRLOWJ}$ is a proper subset of the class $\bf{CS}$. Same results hold true for the class $\bf{GLLOWJ}$.

We recall some results from \cite{owjfa}.
\begin{lemma}\label{regrowj}
$\bf{ROWJ}$ properly includes $\bf{REG}$, i.e., $\bf{REG \subset ROWJ}$.
\end{lemma}

\begin{lemma}\label{cfrowj}
$\bf{CF}$ and $\bf{ROWJ}$ are incomparable.
\end{lemma}
From Lemma \ref{regrowj} and \ref{rowgrlow}, we have the following result.
\begin{proposition}\label{reggrl}
$\bf{REG \subset GRLOWJ}$.
\end{proposition}

\begin{proposition}\label{cfgrl}
The language classes $\bf{CF}$ and $\bf{GRLOWJ}$ are incomparable.
\end{proposition}
\begin{proof}
Being concatenation of two context-free languages, the language $Dc$ $\in$ $\bf{CF}$ but from Lemma \ref{dyckcon}, $Dc \not\in$ $\bf{GRLOWJ}$. Hence, $\bf{CF}$ $\not\subseteq$ $\bf{GRLOWJ}$. From Lemma \ref{cfrowj}, we conclude $\bf{ROWJ \not\subseteq CF}$ and hence, from Lemma \ref{rowgrlow}, we have $\bf{GRLOWJ}$ $\not\subseteq$ $\bf{CF}$.

\end{proof}

\begin{proposition}\label{csgrl}
The language class $\bf{GRLOWJ}$ is a proper subset of the class $\bf{CS}$, i.e., $\bf{GRLOWJ \subset CS}$.
\end{proposition}
\begin{proof}
For a given $\it{GRLOWJFA}$, we give a sketch of the construction of an equivalent Linear bounded
automaton($LBA$).
\begin{enumerate}
    \item[1.] For a given word $w=w_1w_2 \cdots w_n$ from the input alphabet $\Sigma$, the tape of the $LBA$ contains $\$ w_1w_2 \cdots w_n \#$, the tape head is at $w_1$ and the $LBA$ is in the starting state $q_0$.  
    
    \item[2.] Suppose the $LBA$ is in a state $p$. For a rule $(p,x=x_1x_2 \cdots x_i, q)$ of $\it{GRLOWJFA}$, the $LBA$ search for nearest subword $x$ of the word available between the tape head and the right end marker $\#$. By nearest subword we mean that suppose tape head and the right end marker contains $uxv$, then $x$ is not a subword of $u$ and $u'x' \neq x$, where $u'$ is a nonempty suffix of $u$ and $x'$ is a nonempty prefix of $x$.
    
    \item[2.a.] If there is a subword $x$, then the $LBA$ marks it as $X_1X_2 \cdots X_i$.
    
    \item[2.a.a] If there are symbols from $\Sigma$ to the right of $X_i$, then the tape head is at the first symbol right to $X_i$ and the $LBA$ goes to the state $q$.
    
    \item[2.a.b.] If there are no symbols from $\Sigma$ to the right of $X_i$, then the $LBA$ concatenates all the subwords of $w$ that are not marked in capitals such as $X_1X_2 \cdots X_i$. Without loss of generality, we assume the input alphabet contains symbol that are in small letters such as $w_1,w_2$ etc. and for marking it uses symbols that are in capital such as $W_1, W_2$ etc.

    \item[2.b.] If there is no $x$, then the $LBA$ concatenates all the subwords of $w$ that are not marked in capitals such as $X_1X_2 \cdots X_i$.
    
    \item[2.c.] Suppose we get a new word $w'$ from $\Sigma^*$, after concatenation. Then, the $LBA$ adjust the end markers such that there is only $w'$ between end markers, i.e. $\$w'\#$. If the $LBA$ uses 2.a. and 2.a.b., then it goes to the state $q$ and if it uses 2.b., then it stays in the state $p$.
    
    \item[2.c.a.] If $w'= \lambda$, then the $LBA$ accepts the given word $w$ if the $LBA$ is one of the final states of $\it{GRLOWJFA}$. 
    \item[2.c.b.] If $w' \neq \lambda$, then the tape head of the $LBA$ is at the first symbol of $w'$. And it repeats the above steps.
    
\end{enumerate}    

    Hence, we conclude that $\bf{GRLOWJ \subseteq CS}$. But from Lemma \ref{dyckcon}, we have $Dc \not\in $ $\bf{GRLOWJ}$. Hence, $\bf{GRLOWJ \subset CS}$. Similarly, it can be shown that $\bf{GLLOWJ \subset CS}$.

\end{proof}

  \begin{figure}
      \centering

\tikzset{every picture/.style={line width=0.75pt}} 

\begin{tikzpicture}[x=0.75pt,y=0.75pt,yscale=-1,xscale=1]

\draw    (200,127) -- (200,30.03) ;
\draw [shift={(200,28.03)}, rotate = 450] [color={rgb, 255:red, 0; green, 0; blue, 0 }  ][line width=0.75]    (10.93,-3.29) .. controls (6.95,-1.4) and (3.31,-0.3) .. (0,0) .. controls (3.31,0.3) and (6.95,1.4) .. (10.93,3.29)   ;
\draw    (201,240) -- (201,145.41) ;
\draw [shift={(201,143.41)}, rotate = 449.21] [color={rgb, 255:red, 0; green, 0; blue, 0 }  ][line width=0.75]    (10.93,-3.29) .. controls (6.95,-1.4) and (3.31,-0.3) .. (0,0) .. controls (3.31,0.3) and (6.95,1.4) .. (10.93,3.29)   ;
\draw    (353,240) -- (353,145.27) ;
\draw [shift={(353,143.27)}, rotate = 449.3] [color={rgb, 255:red, 0; green, 0; blue, 0 }  ][line width=0.75]    (10.93,-3.29) .. controls (6.95,-1.4) and (3.31,-0.3) .. (0,0) .. controls (3.31,0.3) and (6.95,1.4) .. (10.93,3.29)   ;
\draw    (54.0,240) -- (54.0,145.42) ;
\draw [shift={(54.0,143.42)}, rotate = 450] [color={rgb, 255:red, 0; green, 0; blue, 0 }  ][line width=0.75]    (10.93,-3.29) .. controls (6.95,-1.4) and (3.31,-0.3) .. (0,0) .. controls (3.31,0.3) and (6.95,1.4) .. (10.93,3.29)   ;
\draw    (178,248) -- (78,248) ;
\draw [shift={(76.32,248)}, rotate = 359.27] [color={rgb, 255:red, 0; green, 0; blue, 0 }  ][line width=0.75]    (10.93,-3.29) .. controls (6.95,-1.4) and (3.31,-0.3) .. (0,0) .. controls (3.31,0.3) and (6.95,1.4) .. (10.93,3.29)   ;
\draw    (221,248) -- (321,248) ;
\draw [shift={(323.59,248)}, rotate = 180] [color={rgb, 255:red, 0; green, 0; blue, 0 }  ][line width=0.75]    (10.93,-3.29) .. controls (6.95,-1.4) and (3.31,-0.3) .. (0,0) .. controls (3.31,0.3) and (6.95,1.4) .. (10.93,3.29)   ;
\draw    (55.0,130.0) -- (186.94,27.05) ;
\draw [shift={(188,26)}, rotate = 499.24] [color={rgb, 255:red, 0; green, 0; blue, 0 }  ][line width=0.75]    (10.93,-3.29) .. controls (6.95,-1.4) and (3.31,-0.3) .. (0,0) .. controls (3.31,0.3) and (6.95,1.4) .. (10.93,3.29)   ;
\draw    (349.2,127.22) -- (210,25) ;
\draw [shift={(210.57,25)}, rotate = 400.52] [color={rgb, 255:red, 0; green, 0; blue, 0 }  ][line width=0.75]    (10.93,-3.29) .. controls (6.95,-1.4) and (3.31,-0.3) .. (0,0) .. controls (3.31,0.3) and (6.95,1.4) .. (10.93,3.29)   ;

\draw (187.5,12) node [anchor=north west][inner sep=0.75pt]    {$\bf{CS}$};
\draw (187.5,130) node [anchor=north west][inner sep=0.75pt]    {$\bf{CF}$};
\draw (182.13,242.0) node [anchor=north west][inner sep=0.75pt]    {$\bf{REG}$};
\draw (23.39,130.0) node [anchor=north west][inner sep=0.75pt]    {$\bf{GLLOWJ}$};
\draw (317.03,130.0) node [anchor=north west][inner sep=0.75pt]    {$\bf{GRLOWJ}$};
\draw (327.03,242.0) node [anchor=north west][inner sep=0.75pt]    {$\bf{ROWJ}$};
\draw (27.52,242.0) node [anchor=north west][inner sep=0.75pt]    {$\bf{LOWJ}$};

\end{tikzpicture}

\caption{{\it{GLOWJFA}} language family and Chomsky hierarchy.}
      \label{chlang}
  \end{figure}
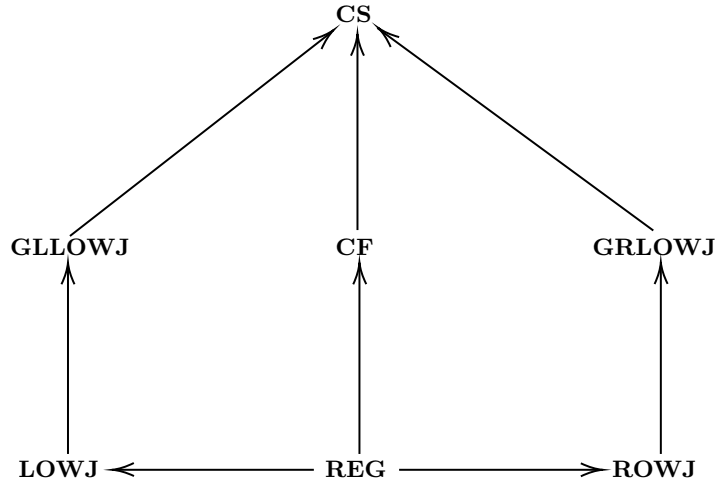  

The results, related  to comparability of language classes of language families $\it{GRLOWJFA/GLLOWJFA}$ and Chomsky hierarchy, presented
in this and above sections are depicted in Fig \ref{chlang}. The arrows represent that one language class is a proper subset of other. If two language classes are not joined by an arrow or a sequence of arrows, then they are incomparable.

\section{Closure Properties}\label{closure}
In this section, we explore closure properties of the language class $\bf{GRLOWJ}$. We show that the class $\bf{GRLOWJ}$ is not closed under intersection, concatenation and reversal. Same results hold true for the class $\bf{GLLOWJ}$.

\begin{note}\label{dyckgrl}
The Dyck language is in the language class $\bf{GRLOWJ}$. An automaton can be constructed similar to the automaton of Example \ref{dyckgll}. 
\end{note}

Now, we give a lemma which we will use to show that the class $\bf{GRLOWJ}$ is not closed under intersection.
\begin{lemma}\label{a^nb^n}
There does not exist any $\it{GRLOWJFA}$ that accepts the language $\{a^nb^n~|~ n \geq 0 \}$.
\end{lemma}
\begin{proof}
If $L=\{a^nb^n~|~ n \geq 0\}$ $\in$ $\bf{GRLOWJ}$, then there exists a $\it{GRLOWJFA}$, $\mathcal{A}=(\Sigma,Q,q_0,F,R)$ such that $L(\mathcal{A})=L.$ Let $l=|Q|$ and $k$ $=$  $max\{|w|~|~(p,w,q)$ $\in R\}$. Consider a natural number $t > lk$. Clearly, $a^tb^t \in L=L(\mathcal{A})$. Then, there exists a sequence of transitions such that $q_0a^tb^t \curvearrowright ^* q_f$, where $q_f \in F$.
\begin{case}
The automaton cannot delete all $a$'s starting from the configuration $q_0a^tb^t$ before deleting a `$b$'. Because in this case the automaton will have to loop to delete all $a$'s and the same loop can be used multiple times to delete more $a$'s than $b$'s and then there would be a mismatch in the indices of `$a$' and `$b$'.
\end{case}
\begin{case}
Since, the automaton cannot delete all $a$'s starting from the configuration $q_0a^tb^t$ before reading a `$b$', let the automaton deletes $t_1$ $a$'s before deleting a `$b$', where $0 \leq t_1 < t$. In this case, we have $q_0a^tb^t \curvearrowright^* q'a^{t-t_1}b^t \curvearrowright ^* q_f$, where $q' \in Q$. The automaton cannot delete all $b$'s starting from the configuration $q'a^{t-t_1}b^t$, same argument would be given as given in case 1. So, let $q_0a^tb^t \curvearrowright^* q'a^{t-t_1}b^t \curvearrowright ^* a^{t-t_1-t_2}q''b^{t-t_3} \curvearrowright q''a^{t-t_1-t_2}b^{t-t_3} \curvearrowright^* q_f$, where $q'' \in Q$, $0 \leq t_2 < k$, $1 \leq t_3 < t$ and $b^i \not\in \Sigma_{q''}$ for $1 \leq i \leq t-t_3$. But in this case, the automaton will also delete the word $a^{t_1+t_2}b^{t_3}a^{t-t_1-t_2}b^{t-t_3}$ using the following sequence of transitions: $q_0a^{t_1+t_2}b^{t_3}a^{t-t_1-t_2}b^{t-t_3} \curvearrowright^* q'a^{t_2}b^{t_3}a^{t-t_1-t_2}b^{t-t_3}$ $\curvearrowright^* q'' a^{t-t_1-t_2}b^{t-t_3}$ $\curvearrowright^*$ $q_f$ and hence $a^{t_1+t_2}b^{t_3}a^{t-t_1-t_2}b^{t-t_3}$ $\in$ $L(\mathcal{A})$. But the word $a^{t_1+t_2}b^{t_3}a^{t-t_1-t_2}b^{t-t_3} \not\in L$, therefore this case is also not possible.
\end{case}

\begin{case}
From the previous two cases, we conclude the automaton cannot delete any `$a$' from the configuration $q_0a^tb^t$. So, the only remaining case is that the automaton will delete $b$'s from the configuration $q_0a^tb^t$ without deleting any `$a$'. Using a similar argument as in case 1, one can show that the automaton cannot delete all $b$'s starting from the configuration $q_0a^tb^t$ before deleting an `$a$'. So, let the automaton deletes first $t_1$ $b$'s with $1 \leq t_1 < t$, i.e., $q_0a^tb^t \curvearrowright^* a^tq'b^{t-t_1} \curvearrowright q'a^tb^{t-t_1} \curvearrowright^* q_f$, where $q' \in Q$ and $b^i \not\in \Sigma_{q'}$ for $1 \leq i \leq t-t_1$. But in this case, the automaton will also delete the word $b^{t_1}a^tb^{t-t_1}$ using the sequence of transitions: $q_0b^{t_1}a^tb^{t-t_1} \curvearrowright^* q'a^tb^{t-t_1} \curvearrowright^* q_f$ and hence, $b^{t_1}a^tb^{t-t_1} \in L(\mathcal{A})$. But, $b^{t_1}a^tb^{t-t_1} \not\in L$, therefore this case is also not possible.

\end{case}
From the above three cases, we conclude $\{a^nb^n~|~ n \geq 0 \} \not\in \bf{GRLOWJ}. $
\end{proof}

\begin{proposition}
$\bf{GRLOWJ}$ is not closed under 
\begin{itemize}
    \item[1.] intersection,
    \item[2.] concatenation,
    \item[3.] reversal.
\end{itemize}
\end{proposition}
\begin{proof}
From Proposition \ref{reggrl}, we have $\bf{REG \subset GRLOWJ}$. Hence, the language $a^*b^*$ $\in$ $\bf{GRLOWJ}$. Also, from Note \ref{dyckgrl}, the Dyck language $D \in \bf{GRLOWJ}$. But, $a^*b^* \cap D= \{a^nb^n~|~ n \geq 0\} \not\in \bf{GRLOWJ}$, by Lemma \ref{a^nb^n}. Hence, $\bf{GRLOWJ}$ is not closed under intersection.

From Lemma \ref{dyckcon}, we have $Dc$ $\not\in$ $\bf{GRLOWJ}$, but $D,\{c\} $ $\in$ $\bf{GRLOWJ}$. Hence, $\bf{GRLOWJ}$ is not closed under concatenation.

An $\it{GRLOWJFA}$ automaton, similar to that of Example \ref{Dc}, can be constructed that accepts the language $cD$ and hence, $cD \in \bf{GRLOWJ}$. We have $(cD)^R=D^Rc$ and $D^R$ is the language obtained by interchanging $a$ and $b$ in $D$. Hence, it can be proved, similar to Lemma \ref{Dc}, that $D^Rc \not\in \bf{GRLOWJ}$. Hence, $\bf{GRLOWJ}$ is not closed under reversal.

\end{proof}

\section{Conclusion}
In this paper, we propose a general one-way jumping automata to understand the discontinuous reading of an input in parts of its substrings. We have defined a new jumping transition model of computation which is names as generalized linear one-way jumping finite automata. It is a generalization of the model one-way jumping finite automata in deleting words, i.e., unlike  $\it{OWJFA}$, the new generalized  model can delete words during computation. We have shown that this new generalization is more powerful than $\it{OWJFA}$. The variants $\it{GRLOWJFA}$ and $\it{GLLOWJFA}$ are compared for their computational power. The language class of $\it{GRLOWJFA}$ has been compared with the language classes of Chomsky hierarchy. Our study on the power of the new generalized model is exhaustive. Closure properties of the language class of $\it{GRLOWJFA}$ have been explored. The study is theoretical in nature.

\end{document}